\documentclass[11pt,aps,prd,showpacs,amsmath,amssymb,floatfix,nofootinbib,superscriptaddress]{revtex4-2}

\usepackage{graphicx}  
\newcounter{subfig}

\usepackage{float} 
\usepackage{booktabs}
\usepackage{makecell}
\usepackage[colorlinks=true,allcolors=BlueViolet]{hyperref}
\usepackage{multirow,bigdelim}
\usepackage[dvipsnames]{xcolor}
\usepackage{ulem}
\usepackage{latexsym}
\usepackage{amsfonts}
\usepackage{amsmath,amssymb}
\usepackage{slashed}
\usepackage{color}
\usepackage{orcidlink}
\usepackage{supertabular} 
\usepackage{placeins}
\usepackage{epsfig}
\usepackage{lipsum}

\newcommand{\ific}{\affiliation{Instituto de F\'isica Corpuscular (centro mixto CSIC-UV), \\
Institutos de Investigaci\'on de Paterna, Apartado 22085, 46071, Valencia, Spain}}

\newcommand{\uv}
{\affiliation{Departamento de F\'{\i}sica Te\'orica and IFIC, Centro Mixto Universidad de Valencia-CSIC Institutos de Investigaci\'on de Paterna, Aptdo.22085, 46071 Valencia, Spain}
}

\begin{document}

\title{Compositeness and decay properties of the flavor partner of $D^{*}_{s0}(2317)$}

\author{J. Sánchez-Illana}
\email{Jesus.Sanchez@ific.uv.es}
\uv

\author{R. Molina}
\email{Raquel.Molina@ific.uv.es}
\uv

\author{Pan-Pan~Shi} \email{Panpan.Shi@ific.uv.es}
\uv\ific

\begin{abstract}
In view of the recent experimental finding of the heavy quark flavor partner of the $D_{s0}(2317)$ in the bottom sector, the $B_{s0}(5700)$, we perform a single and coupled-channel $B^{(*)}\bar{K}^{(*)}$ calculation in the hidden-gauge formalism within the on-shell and off-shell schemes, respectively. By fixing the $B\bar{K}$ bound state mass to the observed state with $J^P=0^+$, we also predict five new states with $J^P=\lbrace 0,1,2\rbrace^+$ in this sector, three of them with $J^P=1^+$. Additionally, in the coupled-channel calculation, two states with masses and decay widths similar to the $B_{sJ}(6114)$ and $B_{sJ}(6063)$ ones, with spins $J=0^+$ and $2^+$, respectivley, and decaying into $B\bar{K}$, are dynamically generated. Moreover, we evaluate the decay widths of the $B_{s0}$ and $B_{s1}$ states generated from $B^{(*)}\bar{K}$ through the isospin breaking decay mode $B_s^0\pi^0$ and $B_s^{*0}\pi^0$, respectively. We find that, because these states have large binding energies, the coupling to the bare component can reduce this decay width by more than half of the decay width, from $50$~keV in the purely molecular model to $10-20$~keV, when the $B_{s0}$ is coupled to the bare component. We also observe that, after considering heavy-quark-spin symmetry violating contributions, the decay width for the $B_{s1}$ becomes smaller than for the $B_{s0}$.
\end{abstract}

\maketitle

\section{Introduction}

The spectrum of mesons in the bottom-strange sector is supposed to be a cleaner scenario for calculations based on the quark model because of several reasons, as the relativistic corrections are smaller than in the two-light meson systems, the relatively small size of the strength of the strong coupling constant and the spin-spin interactions, which become negligible in the heavy quark limit, among other reasons. In spite of this, experimentally, the bottom sector is largely unexplored, with only a few excited mesons observed~\cite{ParticleDataGroup:2024cfk}. In the coming years, there might be a turning point in the spectrum with more experimental information that can be confronted with the theoretical calculations. Indeed, very recently, the LHCb has announced the discovery of the bottom partner of the $D_{s0}(2317)$~\cite{LHCb:2026knl} in the invariant mass distribution of $B_s^0\pi^0$, with a mass of $M_{B_{s0}}= 5698.9\pm 1.5\pm 0.6 $~MeV and a decay width $\Gamma< 9.8 (11.8)$~MeV at $90$\% CL. This observation, if confirmed, challenges the quark model predictions, which usually lie in an energy range $5800-5850$ MeV~\cite{Godfrey:1986wj,Ebert:1997nk,DiPierro:2001dwf,Ebert:2009ua,Sun:2014wea,Godfrey:2016nwn,Hao:2026ppm}. The new meson discovered in the bottom-strange sector is around $75$ MeV below the $B\bar{K}$ threshold. Heavy Meson Chiral Perturbation Theory (HMChPT)~\cite{Burdman:1992gh,Wise:1992hn,Yan:1992gz} successfully describes the $D_{s0}(2317)$ and $D_{s1}(2460)$~\cite{Kolomeitsev:2003ac,Gamermann:2006nm,Faessler:2007gv,Faessler:2007us,Guo:2008gp,Liu:2012zya,Yao:2015qia,Albaladejo:2018mhb,Du:2017zvv,Gil-Dominguez:2023puj} and gives predictions for the $B_{s0}$ of around $5700-5760$ MeV~\cite{Kolomeitsev:2003ac,Guo:2006fu,Guo:2007up,Cleven:2010aw,Cheng:2014bca,Albaladejo:2016ztm,Cheng:2017oqh,Du:2017zvv} (see also~\cite{Bardeen:2003kt,Colangelo:2012xi,Zhang:2024usz}), which are much closer to the experimental observation than the quark model predictions. Among these types of calculations, it is worth highlighting that in ~\cite{Albaladejo:2016ztm}, the mass of the scalar state is predicted to be $m_{B_{s0}}=5711\pm 6$~MeV, from the analysis of LQCD energy levels for $B\bar{K}$ scattering~\cite{Lang:2015hza}, within a framework where the interaction combines the LO HMChPT with the coupling to a bare component. A work in the same line obtains a mass $m_{B_{s0}}=5730^{+2.4}_{-1.5}$~MeV by using Hamiltonian Effective Field Theory~\cite{Yang:2022vdb}. An important test for these approaches, where the heavy-light meson loops play a crucial role to explain the masses of these excited states, are the radiative and hadronic decays. See~\cite{Hao:2026ppm}, that incorporates the two-meson coupled-channel dynamics into the quark model and studies the radiative decays for the $B_{s0}$ and $B_{s1}$. The hadronic isospin violating decay mode is supposed to be at least one order of magnitude smaller for the purely CQM states~\cite{DiPierro:2001dwf,Godfrey:2016nwn,Han:2023wqq,Colangelo:2003vg,Colangelo:2005hv} than in the case where the state couples significantly to the two-meson component~\cite{Fu:2021wde,Gil-Dominguez:2023puj,Su:2025aiz}. In~\cite{Zhou:2026vjp} a mixed state for the the $D_{s0}(2317)$ is considered and the decay to $D_s^+\pi^0$ is evaluated. Lately, in~\cite{Albaladejo:2026yfz}, the invariant mass distribution of the $B_{s0}$ state observed by the LHCb in the $B_s^0\pi^0$ channel has been analyzed within HMChPT with the input by LQCD~\cite{Albaladejo:2016ztm}, the decay width of the $B_{s0}$ is obtained within the range of $30-100$~keV for the sets of models used, and the mass of the $B_{s1}$ being predicted to be about $5740-5780$~MeV.

On the other hand, we can also find in the literature several results from LQCD simulations. Earlier calculations gave a mass for the $B_{s0}$ state around $25$~MeV below the $B\bar{K}$ threshold~\cite{Gregory:2010gm}. More recent refined LQCD simulations give binding energies closer to the experimental one within the range $60-80$~MeV~\cite{Lang:2015hza,Hudspith:2023loy,Gayer:2024akw,Hudspith:2026vxt,Guyton:2026mja}. Notably, in~\cite{Lang:2015hza}, the inclusion of $B\bar{K}$ operators is considered and a binding energy of about $60$~MeV is obtained. Lately, the $B_s$ spectrum has also been studied in~\cite{Gayer:2024akw} for $m_\pi=391$ MeV providing a value of $79$~MeV for the scalar state binding energy. However, in the simulation, the authors utilized the single-meson operators, while the $B\bar K$-type interpolating operators were excluded. In the recent work~\cite{Guyton:2026mja}, $B^{(*)}\bar{K}$ operators are also considered, the properties of the $B_{s0}$ and $B_{s1}$ are studied, and after applying the Weinberg compositeness criterion~\cite{Weinberg:1965zz}, the authors conclude that these states are predominantly molecular. 

In the bottom sector, two excited states already were discovered before the $B_{s0}$ observation, the $B_{sJ}(6063)$ and the $B_{sJ}(6114)$~\cite{LHCb:2020pet}, whose splitting cannot be accommodated within the quark-model predictions~\cite{Godfrey:1986wj,Ebert:1997nk,DiPierro:2001dwf,Ebert:2009ua,Sun:2014wea,Godfrey:2016nwn}. These, depending on whether they decay to $B\bar{K}$ or $B^*\bar{K}$ have masses and decay widths which are,
\begin{eqnarray}
   && B_{SJ}(6063): m=6063.5\pm 1.2~(6108.8\pm 1.1)~\mathrm{MeV}, \quad \Gamma=26\pm 4~(22\pm 5)~\mathrm{MeV}\nonumber\\
   &&B_{SJ}(6114): m=6114\pm 3~(6158\pm 4)~\mathrm{MeV},\quad \Gamma=66\pm 18 ~(72\pm 18)~\mathrm{MeV}\ ,
\end{eqnarray}
where the numbers in parentheses correspond to the $B^*\bar{K}$ decay~\cite{LHCb:2020pet}. In~\cite{Sanchez-Illana:2026cyv}, a new prediction for the $B_{s0}$ mass is obtained based on the assumption that these two states are $B\bar{K}^*$ and $B^*\bar{K}^*$ molecular states decaying into $B^*\bar{K}$ within the hidden-gauge formalism, predicting a value of $5760$
MeV for the mass of the $B_{s0}$ state bound in $B\bar{K}$. Predictions from unitarized EFT's and other phenomenological approaches depend on how the degrees of freedom, cutoff, LECs, or the parameters related to the bare component, if included, are constrained. The evaluation of the isospin-violating hadronic decays, highly dependent on the size of the coupling of the exotic state to the two-meson, can provide valuable information on the properties of these states, and suppose a validation for the theoretical framework. In this work, first, we revisit our predictions on the $B^{(*)}\bar{K}^{(*)}$ states, which are now based on the experimental observation of the $B_{s0}$ state. Second, we evaluate the isospin-breaking decay of the $B_{s0}(5700)$ to $B_s^0\pi^0$ within HMChPT and investigate how the size of this decay depends on the compositeness of the state, by considering a mixed state with the coupling of the two-meson component to the $\bar{b}s$ bare state. This article continues as follows. In Section II we briefly describe how the scattering amplitude for the heavy-light meson interaction is constructed and the formalism used for the evaluation of the isospin-breaking decay mode of the $B_{s0}(5700)$. We also calculate this decay mode for the case of the $B_{s1}$ state. In Section III we present and discuss our results, which are presented for single and coupled channels. Finally, we give some conclusions in Section IV.

\section{Formalism}\label{Sec:formula}

\subsection{Interaction in the $B^{(*)} \bar{K}^{(*)}$ system}~\label{se:int}
Below, we briefly summarize the scattering in the $B^{(*)}\bar{ K}^{(*)}$ systems. The $B\bar{K}$ and $B^* \bar{K}$ interactions are related to the $B_{s0}$ and $B_{s1}$ states. The interactions in the systems studied here are encoded within the framework of the local hidden gauge symmetry~\cite{Nagahiro:2008cv,Molina:2009ct,Molina:2010tx}. Following our previous study~\cite{Sanchez-Illana:2026cyv}, these interactions are mediated by the exchange of the $\rho$ and $\omega$ mesons. The corresponding potentials for $B\bar{K}$ and $B^* \bar{K}$ are given by
\begin{align}
V_{B\bar{K}}(s)&=\frac{g^2}{2}\left[3D_{su}(m_\rho)+D_{su}(m_\omega)\right],\label{eq:rho}\\
V_{B^*\bar{K}}(s)&=-\frac{g^2}{2}\left[3D_{su}(m_\rho)+D_{su}(m_\omega)\right]\,\varepsilon(p_1)\!\cdot\!\varepsilon^\ast(k_1),
\label{eq:s1}\end{align}
respectively, where $\varepsilon(p_1)$ and $\varepsilon^\ast(k_1)$ denote the polarization vectors for the initial and final states, respectively. The coupling constant is $g=m_{\rho}/2F_{\pi}$ with $F_{\pi}=93$~MeV the pion decay constant and $m_{\rho}$ the $\rho$ meson mass. Here, the function $D_{su}(m)$ is defined as
\begin{align}
D_{su}(m)=\frac{s-u}{t-m^2},
\label{eq:fullvmp}
\end{align}
where $s=(p_1+p_2)^2,~t=(p_1-k_1)^2$, and $u=(p_1-k_2)^2$ are the Mandelstam variables in the reaction $B^{(*)}(p_1)\bar{K}(p_2)\to B^{(*)}(k_1)\bar{K}(k_2)$. The $B\bar{K}^*$ and $B^*\bar{K}^*$ potentials for $\rho$ and $\omega$ exchange are evaluated in a similar way. These are given in Table I of~\cite{Sanchez-Illana:2026cyv} for the possible spins and in coupled-channels also. In~\cite{Sanchez-Illana:2026cyv} the partial wave projection is explained. As a novelty from our previous work, for the vector-vector interaction, i.e., in the $B^*\bar{K}^*$ system, we include a contact term, which is given by~\cite{Molina:2010tx}
\begin{equation}
    \mathcal{L}_{4V}=\frac{1}{2}g^2\langle\left[V_\mu,V_\nu\right]V^\mu V^\nu\rangle.
    \label{Eq:ctermL}
\end{equation}
This term produces a mass difference between the resonances generated by the rescattering of the isoscalar $B^*\bar{K}^*$ system with different total angular momentum. Hence, the corresponding Feynman diagrams in this case are depicted in Fig.~\ref{Fig:treelevels}. The contact terms are listed in Table~\ref{Tab:ctermcoef} for different spins in ~\cite{Molina:2010tx}\footnote{The flavor coefficients arising from the $4V$ contact interaction in the $B^*\bar K^*$ system are identical to those obtained in the $D^*K^*$ sector, which are given in Table 13 of \cite{Molina:2010tx}. This correspondence originates from the replacement $B^*\to \bar{D}^*$ and the fact that the strong interaction is invariant under charge conjugation symmetry.}, 
and we can write, $V(s)=V_{ct}+V_{B^{*}\bar{K}^{*}}(s)$, for the vector-vector interaction, summing the contact and vector-meson exchange contributions.

\begin{figure}[h]
    \centering
    \includegraphics[height=0.25\columnwidth]{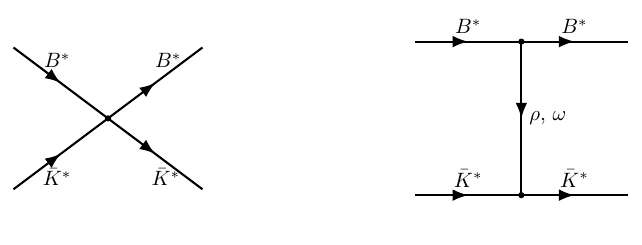}
    \caption{Feynman diagrams included in the potential calculations for $B^\ast\bar K^\ast$ system.}
    \label{Fig:treelevels}
\end{figure}

\begin{table}[htbp]
\centering
\caption{Contact term for $B^\ast\bar K^\ast$ with $I=0$ and $J=0,1,2$ systems.}
\label{Tab:ctermcoef}
\renewcommand{\arraystretch}{1.2}
\begin{tabular}{lcc}
\hline\hline
Channel & $J^P$ & $V_{ct}$ \\
\hline
$B^\ast\bar K^\ast$ & $0^+$ & $4g^2$  \\
$B^\ast\bar K^\ast$ & $1^+$ & $6g^2$  \\
$B^\ast\bar K^\ast$ & $2^+$ & $-2g^2$ \\
\hline\hline
\end{tabular}
\end{table}

The scattering of the $B^{(*)}\bar{K}^{(*)}$ system is described by the  Lippmann-Schwinger (LS) equation, that in the on-shell scheme can be reduced to 
\begin{align}
T(s)&=\left[V(s)^{-1}-G(s)\right]^{-1}=\left[1-V(s)G(s)\right]^{-1}V(s).
\end{align}
where the total effective potential is, $V(s)=V_{B^{(*)}\bar{K}^{(*)}}(s)$, for the two-pseudoscalar and pseudoscalar-vector interaction. 

The two-point loop function $G(s)$ is regularized by the hard cutoff $\Lambda$
\begin{align}
G(s)=\int_{\Lambda}\frac{d^3{\boldsymbol q}}{(2\pi)^3}\frac{\omega_1(\boldsymbol q)+\omega_2(\boldsymbol q)}{2\omega_1(\boldsymbol q)\omega_2(\boldsymbol q)}\frac{1}{s-\left[\omega_1(\boldsymbol q)+\omega_2(\boldsymbol q)\right]^2+i\epsilon},
\label{Eq:green}
\end{align}
with $\omega_{\alpha}(\boldsymbol q)=\sqrt{m_{\alpha}^2+\boldsymbol q^2}$.

The possible structures of the $B^{(*)}\bar{K}^{(*)}$ system, including the bound state, virtual state, and resonance, are associated with the pole positions of the $T$-matrix. The pole positions $s_R$ are determined by $\text{Det}\,[1-V(s_R)G(s_R)]=0$ in the complex-$s$ plane. According to the analytic continuation, the complex-$s$ plane in the $N$-coupled-channel system can be divided into the $2^N$ Riemann sheets (RSs), labeled by $(\pm,...,\pm)$ in terms of the sign of the imaginary part of the two-point functions.
The physical RS corresponds to $(+,...,+)$. For the $i$-th channel, the unphysical RS can connect to the physical one through the relation
\begin{align}
G_i^{-}(s)=G_i^{+}(s) +\frac{ip_i}{4\pi \sqrt{s}},
\end{align}
where $p_i$ is the three-momentum in the center-of-mass (c.m.) frame, and the superscripts $+$ and $-$ denote the two-point function on the physical and unphysical RSs, respectively. We perform here on-shell single channel and off-shell coupled channel calculations. Regarding the latter, we refer to~\cite{Sanchez-Illana:2026cyv} for a detailed description of the formalism. The modification in the interaction included in the present work is the four-vector contact term discussed.

The effective coupling of the pole to the $i$-th channel is extracted from the residue of the $T$-matrix
\begin{align}
g^2_{ij}=\lim_{s\to s_R}(s-s_R)T_{ij}(s),
\end{align}
where $T(s)$ is located on the RS where the pole appears. For a bound state, the compositeness $X_i$ is evaluated from~\cite{Gamermann:2009uq,Hyodo:2011qc,Guo:2015daa}
\begin{align}
X_i= -g_i^2\left[\frac{\partial G_i(s)}{\partial s}\right]_{s=s_R}.
\label{Eq:posibility-coupled}
\end{align}
We further characterize the near-threshold interaction through the $S$-wave scattering length $a_0$ and effective range $r_0$, which are extracted from the single-channel $T$-matrix
\begin{align}
\frac{1}{a_0}=&\left[\frac{-8\pi \sqrt{s}}{T(s)}\right]_{s=m_{\text{thr.}}^2}\\
r_0=&2\left[\frac{\partial}{\partial p^2}\left(\frac{-8\pi \sqrt{s}}{ T(s)}+ip\right)\right]_{s=m_{\text{thr.}}^2},
\end{align}
with $m_{\text{thr.}}=m_{B^{(*)}}+m_{\bar{K}^{(*)}}$.

\subsection{Hadronic decay of the $B_{s0}$ and $B_{s1}$ to $B_s^{(*)}\pi$ }~\label{se:hade}
The hadronic decays of the $B_{s0}$ and $B_{s1}$ states can occur through the triangle mechanism. As illustrated in Fig.~\ref{Fig:decay_diam}, we consider the $K^*$ and $B^*$ exchange contributions to the decay $B_{s0}\to B_s\pi^0$, while the $K^*$ and $B$ exchange contributions are included for $B_{s1}\to B^*\pi^0$. The $B^*$ exchange for the latter decay mode is neglected since, in the triangle diagrams of Fig.~\ref{Fig:decay_diam}, the vertex of $B^*\to B^* \pi$ is suppressed by ${\cal O}(p^2)$ compared with that of $B^*\to B\pi$~\cite{Ruiz-Femenia:2003jdx,Oset:2022xji}. Besides, owing to the $\pi^0-\eta$ mixing, we also consider the decay process $B_{s0(1)}\to B_s^{(*)}\eta \to B_s^{(*)}\pi^0$~\cite{Cleven:2014oka,Fu:2021wde,Gil-Dominguez:2023puj}.

With explicit the vertices derived from the local hidden gauge symmetry~\cite {Sanchez-Illana:2026cyv}, the triangle amplitudes depicted in Fig.~\ref{Fig:decay_diam} with the momenta assignment in Fig. \ref{Fig:momentum_diam} are\footnote{In the expression below and in what follows we neglect terms like $\varepsilon(B^*_{(s)})\cdot k(B^*_{(s)})$ for heavy vector mesons, that involve negligible factors of the order ${\cal O}(\vert \vec{k}(B^*_{(s)})\vert/m_{B^*_{(s)}})$, with $k$ the four-momentum of the vector meson. }
\begin{align}
-i{t}_{B_{s0}}=&ig_{B_{s0},B\bar{K}}g^2c_1\left(\frac{(p+q-p_f)\cdot (p-q+p_f)}{[q^2-m_{B}^2+i\varepsilon][(p-q)^2-m_K^2+i\varepsilon][(q-p+p_f)^2-m_{K^*}^2+i\varepsilon]}\right.\nonumber\\
&+\left.\frac{(q+p_f)\cdot (2p-q-p_f)}{[q^2-m_{B}^2+i\varepsilon][(q-p_f)^2-m_{B^*}^2+i\varepsilon][(p-q)^2-m_{K}^2+i\varepsilon]}\right),\nonumber\\
-i{t}_{B_{s1}}=&ig_{B_{s1},B^*\bar{K}}g^2c_1\left(\frac{(p+q-p_f)\cdot (p-q+p_f)g_{\mu\nu}
}{[q^2-m_{B^*}^2+i\varepsilon][(p-q)^2-m_K^2+i\varepsilon][(q-p+p_f)^2-m_{K^*}^2+i\varepsilon]}\right.\nonumber\\
&-\left.\frac{(2\,p_{f\,\mu}-q_{\mu})(p_{\nu}+p_{f\,{\nu}}-2\,q_{\nu})}{[q^2-m_{B^*}^2+i\varepsilon][(q-p_f)^2-m_{B}^2+i\varepsilon][(p-q)^2-m_{K}^2+i\varepsilon]}\right)\varepsilon^\mu \varepsilon^{\prime\nu\ast}.
\label{Eq:amplitude}
\end{align} 
The constant $c_1$ is listed in Table~\ref{Tab:coefficients} for each diagram in Fig.~\ref{Fig:decay_diam}. 
The mixing angle $\tilde \epsilon$ in Table~\ref{Tab:coefficients} is given by 
\begin{align} 
\tilde \epsilon=\frac{\sqrt 3}{4}\frac{m_d-m_u}{m_s-\hat m},
\end{align}
with $\hat m=(m_u+m_d)/2$ the average mass of the $u$ and $d$ quarks.

\begin{table}[htbp]
\caption{Masses and coefficients of the diagrams in Fig.~\ref{Fig:decay_diam} that contribute to the $B_{s0}\to B_s\pi^0/B_{s1}\to B_s^*\pi^0$ decay. $m_1$, $m_2$, $M$ denote the mass of the particles inside the loop, being $M$ the mass of the exchanged particle between the heavy ($m_1$) and the light ($m_2$) one, while $m_{1,2f}$ stand for the masses of the particles in the final state (as in Fig. 19 of~\cite{Gil-Dominguez:2023puj}).}
\label{Tab:coefficients}
\renewcommand{\arraystretch}{1.2}
\begin{tabular*}{\columnwidth}{@{\extracolsep\fill}lccccc}
\hline\hline 
\hline $c_1$ & $m_1$ & $m_2$ & $M$ & $m_{1 f}$ & $m_{2 f}$ \\
\hline $1 / \sqrt{2}$ & $B^{0}/B^{*0}$ & $\bar K^0$ & $K^{*0}$ & $B_s/B_s^{*}$ & $\pi^0$ \\
\hline $-1 / \sqrt{2}$ & $B^{+}/B^{*+}$ & $K^{-}$ & $K^{*+}$ & $B_s/B_s^{*}$ & $\pi^0$ \\
\hline $1 / \sqrt{2}$ & $B^{0}/B^{*0}$ & $\bar K^0$ & $B^{*0}/B^0$ & $\pi^0$ & $B_s/B_s^{*}$ \\
\hline $-1 / \sqrt{2}$ & $B^{+}/B^{*+}$ & $K^{-}$ & $B^{*+}/B^+$ & $\pi^0$ & $B_s/B_s^{*}$ \\
\hline $-\tilde{\epsilon} \sqrt{\frac{3}{2}}$ & $B^{0}/B^{*0}$ & $\bar K^0$ & ${K}^{*0}$ & $B_s/B_s^{*}$ & $\eta$ \\
\hline $-\tilde{\epsilon} \sqrt{\frac{3}{2}}$ & $B^{+}/B^{*+}$ & $K^{-}$ & $K^{*+}$ & $B_s/B_s^{*}$ & $\eta$ \\
\hline $-\tilde{\epsilon} \frac{1}{\sqrt{6}}$ & $B^{0}/B^{*0}$ & $\bar K^0$ & $B^{*0}/B^0$ & $\eta$ & $B_s/B_s^{*}$ \\
\hline $-\tilde{\epsilon} \frac{1}{\sqrt{6}}$ & $B^{+}/B^{*+}$ & $K^{-}$ & $B^{*+}/B^{+}$ & $\eta$ & $B_s/B_s^{*}$ \\
\hline\hline
\end{tabular*}
\end{table}

For the evaluation of the triangle loops for $B_{s0}\to B_{s}\pi^0$, it is convenient to perform an integration over the $q^0$-component of the loop momentum analytically. After integrating the $q^0$-component using the Residue Theorem, the amplitude is reduced to~\cite{Gil-Dominguez:2023puj}
\begin{align}
-i{t}_{B_{s0}}=\frac{ig_{B_{s0},B\bar{K}}g^2c_1}{32 \pi^2} \left[I_1(\omega_{B},\omega_{\bar K}, \omega_{K^*})+I_1(\omega_{B}, \left.\omega_{\bar K}, \omega_{B^*})\right|_{p_f^0\to p^0-p_f^0, \vec{p}_f\to-\vec{p}_f}\right],
\end{align}
where the energy of the hadron $i$ is $\omega_i=\sqrt{m_i^2+\vec{p}_i^{\,2}}$ with the corresponding three-momentum $\vec{p}_i$, and $I_1(\omega_1, \omega_2, \omega_M)$ is defined in Eq.~\eqref{AppI1}. The amplitude for $B_{s1}\to B_s^* \pi^0$ is discussed in Appendix~\ref{AppA}.

\begin{figure}[h]   
\includegraphics[height=0.25\columnwidth]{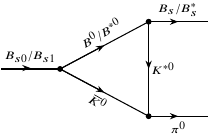} \hspace{1.0cm} 
\includegraphics[height=0.25\columnwidth]{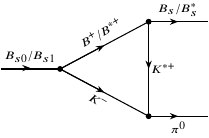}\\
\includegraphics[height=0.25\columnwidth]{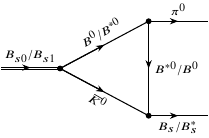} \hspace{1.0cm}
\includegraphics[height=0.25\columnwidth]{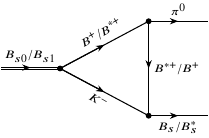}
   \caption{Feynman diagrams for the decay $ B_{s0}\to B_{s}\pi^0$ and $ B_{s1}\to B^*_{s}\pi^0$.}
    \label{Fig:decay_diam}
\end{figure}

\begin{figure}[h]
    \centering
    \includegraphics[height=0.24\columnwidth]{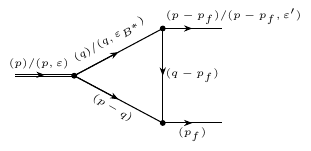}
    \hfill
    \includegraphics[height=0.24\columnwidth]{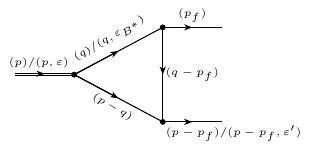}
    \caption{Feynman diagrams of the assignment of momenta. The assignment for the two top diagrams in Fig.~\ref{Fig:decay_diam} on the left and the two bottom diagrams in Fig.~\ref{Fig:decay_diam} on the right.}
    \label{Fig:momentum_diam}
\end{figure}

The final decay widths for the processes $B_{s0}\to B_s\pi^0$ and $B_{s1}\to B_s^*\pi^0$ are obtained from the corresponding amplitudes according to
\begin{align}
\Gamma(B_{sJ})=\frac{1}{2J+1}\int\frac{\text{d}\Omega}{32\pi^2}\frac{|\vec p_f|}{m_{B_{s0(s1)}}^2}\sum_{\lambda_i,\lambda_f} {\left|t_{B_{sJ}}\right|^2},
\end{align}
where $p_f$ denotes the momentum of the final states in the c.m. frame, and $J=0$ and $1$ for the decays of $B_{s0}$ and $B_{s1}$ mesons, respectively. Here $\lambda_i$ and $\lambda_f$ label the polarization components for the initial and final hadrons. 

\subsubsection{Consideration of the bare component in the $B_{s0}$ and $B_{s1}$ decay width calculation}

We also consider the impact of the inclusion of the coupling to a bare component for the $B^{(*)}\bar{K}$ states in the decay width to $B_s^{(*)}\pi^0$. To that end, we modify the one-channel on-shell calculation described in Sec.~\ref{se:int} by adding the bare component to the exchange term. This is, the interaction is now given by,
\begin{eqnarray}
    V'_{B^{(*)}\bar{K}}=V_{B^{(*)}\bar{K}}+V_{bare}
\end{eqnarray}
with $V_{B^{(*)}\bar{K}}$ the vector-meson exchange interaction given in Eqs.~(\ref{eq:rho}) and (\ref{eq:s1}). 
The vertex for the coupling of the $B^{(*)}\bar{ K}$ state to the bare $b\bar s$ state is incorporated through the chiral perturbative theory with heavy quark spin symmetry~\cite {Falk:1991nq,Albaladejo:2016ztm}. At leading order (LO), the effective potential is~\cite{Albaladejo:2016ztm,Albaladejo:2018mhb,Gil-Dominguez:2023puj}
\begin{align}
V_{bare}(s)=\frac{V^2_{b \bar{s}}(s)}{s-m_{b\bar s}^2},
\end{align}
where $s=E^2$, the squared energy in the center-of-mass frame. The effective coupling $V_{b \bar{s}}(s)$ is
\begin{align}
V_{b \bar{s}}(s)=-\frac{c}{F_\pi} \sqrt{M_{B^{(*)}} m_{b \bar{s}}} \,\frac{s+m_{\bar{K}}^2-M_{B^{(*)}}^2}{\sqrt{s}},
\end{align}
where $m_{b \bar{s}}$ is the bare mass of $J^P=0^+$ and $1^+$ $b\bar s$ state, and $c=0.74\pm 0.05$ is determined from a reanalysis of LQCD data~\cite{Albaladejo:2016ztm}. The LQCD calculation was performed at a pion mass of $m_\pi\simeq156(7)(2)~\mathrm{MeV}$~\cite{Lang:2015hza}. Since this value is already close to the physical pion mass, we neglect the residual light-quark-mass dependence of the coupling $c$. The bare masses $m_{b\bar s}$ are taken from unquenched quark-model calculations~\cite{Ortegaetal,Luetal,Lakhinaetal,DiPierroetal,Godfreyetal2016,Liuetal,Sunetal,Godfreyetal1991,Ebertetal}. For the different values of the bare masses given in the literature, we evaluate the hadronic decay of the $B_{s0(1)}$ into $B_s^{(*)}\pi^0$ as described in Sec.~\ref{se:hade}. The cutoff is varied to reproduce the experimental value of the mass of the $B_{s0}$. Thus, the decay width and scattering parameters obtained depend on the compositeness of the state. 
\section{Numerical Result}\label{sec:result}
In this section, first, we present the results for the $B^{(*)}\bar{K}^{(*)}$ interaction in the on-shell single channel calculation, updated to reproduce the mass of the new $B_{s0}$ state discovered by the LHCb~\cite{LHCb:2026knl}. Then, we study the isospin-breaking decay mechanism, $B_{sJ}\to B_s^{(*)}\pi^0$, and we investigate how this decay width, the scattering length and the effective range, vary with the compositeness. Finally, we show the results for the off-shell coupled-channel calculation.
\subsection{Single-channel on-shell calculation of the $B^{(*)}\bar{K}^{(*)}$ system}
In this section we perform a single-channel calculation for the $B^{(*)}\bar{K}^{(*)}$ interaction. The results are given in Table~\ref{Tab:onshpole_Ierr}. A hard cutoff has been used in this case. The coupling to the bare component is omitted here. While in~\cite{Sanchez-Illana:2026cyv}, binding energies of around $20$~MeV were obtained, now these are larger, of $75$~MeV for the $B_{s0(1)}$ and about $170-200$~MeV for the bound states obtained in the $B^{*}\bar{K}^*$ system. Note that, the $J^P=2^+$ state is now slightly more bound than the ones obtained for $J^P=0^+, 1^+$ in the case of the $B^{*}\bar{K}^*$ system, while these states were degenerate in our previous calculation~\cite{Sanchez-Illana:2026cyv}. This degeneracy is now broken due to the inclusion of the four-vector contact term. In total, six bound states are obtained in the calculation because of the large attraction between the bottom and strange mesons. 
In this calculation, $t$ in the vector-meson propagator of Eqs.~(\ref{eq:rho}) and (\ref{eq:s1}), has been neglected after $s$-wave projection, as in~\cite{Molina:2010tx}. We can see that if one reproduces well the new $B_{s0}$ state, only one of the two states considered in~\cite{Sanchez-Illana:2026cyv} could also be described reasonably well, where the $B_{sJ}(6063)$~\cite{LHCb:2020pet} could correspond to our $B^{*}\bar{K}^*$ with $J^P=0^+$\footnote{But now decaying into $B\bar{K}$ instead of $B^*\bar{K}$.}.
The spin partners of the $B_{s0}$ are now predicted. For the $B_{s1}$ with the lowest mass coupling strongly to $B^*\bar{K}$ we obtain that its mass should be about $5744$~MeV. Our result is compatible with the joint analysis of LQCD and experimental data~\cite{Albaladejo:2026yfz}. Apart from this state, other two states with $J^P=1^+$ are predicted, coupling strongly to $B\bar{K}^*$ and $B^*\bar{K}^*$, with masses $5982$~MeV and $6056$~MeV, respectively.
\begin{table}[htbp]
\caption{{On-shell single channel results using the $B_{s0}(5700)$ as input. We implement errors from $\Lambda=874\pm8$ MeV.}}
\label{Tab:onshpole_Ierr}
\renewcommand{\arraystretch}{1.2}
\begin{tabular*}{\columnwidth}{@{\extracolsep\fill}lccccc}
\hline\hline 
 $J^{P}$ &  Channel & Pole [MeV] & RS  & Thr. [MeV] & $E_b$ [MeV]\\[3pt]     
\hline
$0^{+}$  & $ B\bar K$ & $5698.9\pm1.5$
     & $(+)$  & $5775.0$ &$-76.1\pm1.5$\\[3pt]
\hline
$1^{+}$  & $ B^\ast\bar K$ & $5744.0\pm1.5$
     & $(+)$ & $5820.0$ &$-76.0\pm1.5$\\[3pt]
\hline     
$1^{+}$  & $ B\bar K^\ast$ & $5981.9\pm1.4$
     & $(+)$ & $6173.0$ &$-191.1\pm1.4$\\[3pt]
\hline
%
$0^{+}$  & $ B^\ast\bar K^\ast$ & $6046.6\pm1.5$
     & $(+)$  &  $6218.0$ & $-171.4\pm1.5$\\[3pt]
     \hline
$1^+$  & $ B^\ast\bar K^\ast$ & $6056.4\pm1.5$
     & $(+)$  &  $6218.0$ & $-161.6\pm1.5$\\[3pt]
     \hline
$2^+$  & $ B^\ast\bar K^\ast$ & $6017.0\pm1.5$
     & $(+)$  &  $6218.0$ & $-201.0\pm1.5$\\[3pt]
\hline\hline
\end{tabular*}
\end{table}
\subsection{Decay width of the $B^{(*)}\bar{K}$ states to $B^{(*)}_{s}\pi^0$}

\begin{figure}[h]
    \centering
    \includegraphics[height=0.35\columnwidth]{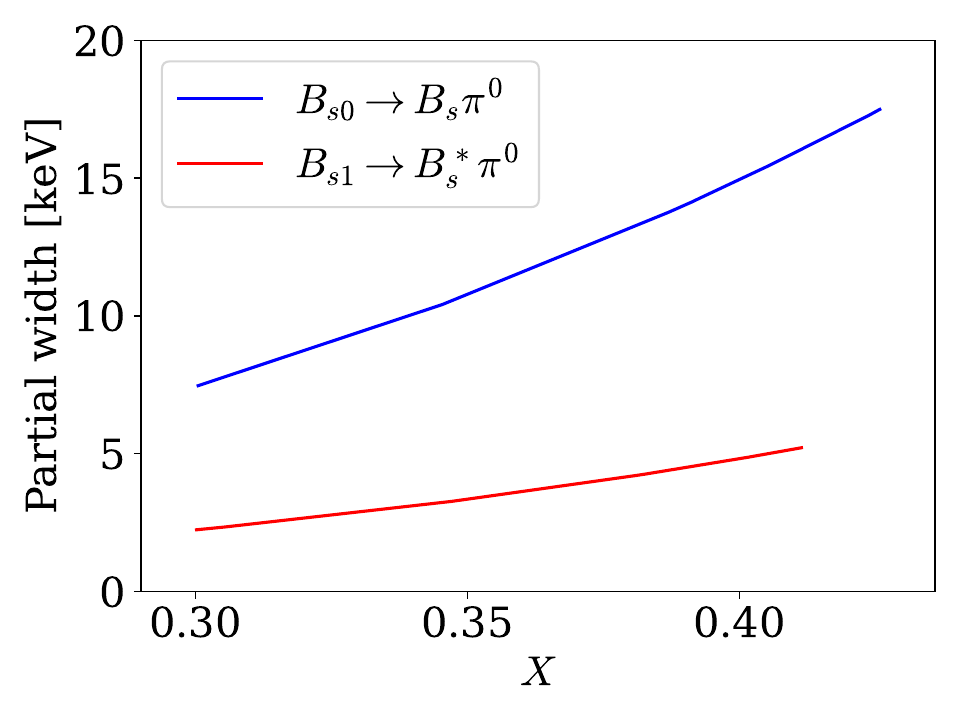}
    \caption{Partial widths as a function of compositeness $X$ for the processes $B_{s0}\to B_s \pi^0$ and $B_{s1}\to B_s^*\pi^0$. The decay widths are obtained from Table~\ref{tab:bare_cutoff1} and Table~\ref{tab:bare2_cutoff1}.}
    \label{Fig:partial_width}
\end{figure}

Here, we evaluate the decay width of the $B_{s0}$ and $B_{s1}$ states, coupling to $B^{(*)}\bar{K}$, to $B_{s}^0\pi^0$ and $B_{s}^{*0}\pi^0$, respectively. Additionally, we determine how this and other observables vary with the compositeness. Firstly, we utilize the calculation of the decay width as explained in the previous section with no bare component, i. e., $c=0$. We obtain, 
\begin{eqnarray}
    \Gamma(B_{s0}\to B_s^0\pi^0)= (53\pm10) ~\text{keV};\quad 
    \Gamma(B_{s1}\to B_s^{*0}\pi^0)= (21\pm4) ~\text{keV}\ .\label{eq:dec}
\end{eqnarray}
If we remove the two bottom diagrams in Fig.~\ref{Fig:decay_diam}, the decay widths of the $B_{s0}$ and $B_{s1}$ become similar to each other, about $22$ keV, consistently with HQSS. However, these two bottom diagrams involve the exchange of heavy mesons, providing heavy quark spin violating contributions to the decay widths, hence, explaining the difference between them. Secondly, we now include the bare component, by taking the value of $c$ fixed as determined in previous analyses~\cite{Albaladejo:2016ztm,Albaladejo:2026yfz}, $c=0.74\pm 0.05$. The cutoff parameter is now varied to reproduce the $B_{s0}$. The results are given in Tables~\ref{tab:bare_cutoff1} and \ref{tab:bare2_cutoff1}, for the $B_{s0}$ and $B_{s1}$, respectively. When varying the cutoff to reproduce the $B_{s0}$ state for the different values of the bare components, the cutoff is in the range $\Lambda\in[300, 450]$~MeV. If instead, a Gaussian regulator is used in the regularization of the loop integral~\cite{Albaladejo:2016ztm}, we obtain larger values of $\Lambda\in[500,800]$, consistent with the cutoff in Ref.~\cite{Albaladejo:2016ztm}. The result for this calculation is in the Appendix \ref{AppB}, and we can see that the values of the decay widths and scattering parameters obtained are very similar to the ones presented here. This means that our results are quite stable against the regularization procedure used. For the values of the bare masses studied, the scattering length takes values in the range of $[-0.9,-0.86]$~fm, and the effective ranges are in the interval $[0.25,0.38]$~fm. We observe that the decay widths of the $B_{s0}$ and $B_{s1}$ are substantially reduced due to consideration of the bare component. For reasonable values of the bare component, the decay width has dropped from $50$~keV to $10-20$~keV in the case of the $B_{s0}$. Since the binding energy of the $B_{s0}$ is not small, we could expect that this decay width can be smaller than the predictions from the purely molecular model. It is clear that the experimental data do not have enough precision to measure this tiny decay width, since at present the upper limit is set to $\Gamma  <9.8(11.8)$~MeV~\cite{LHCb:2026knl} and the analysis of the data can carry on large systematic uncertainties providing decay widths of around $30-100$~keV for the sets of models used in~\cite{Albaladejo:2026yfz}. 
Still, as it is shown in Fig.~\ref{Fig:partial_width}, the partial decay widths are sensitive to the variation of the compositeness both for the $B_{s0}$ and $B_{s1}$ decays. Therefore, the future precise measurement of this decay width can be very useful to understand the nature of these state.

\begin{table}[htbp] 
\centering 
\caption{Values of the bare mass $M_{\rm bare}$, the corresponding scattering length $a_0$, effective range $r_0$, compositeness $X$, complementary component $1-X$, pole coupling $g_{\rm pole}$, and decay width $\Gamma(B_{s0})$.
} \label{tab:bare_cutoff1}
\begin{tabular}{cccccccc} 
\hline
Reference & $M_{\rm bare}$ [MeV]  &$a_0$ [fm] &$r_0$ [fm] &$1-X$ &$X$ &$g_{\rm pole}$ [MeV]&$\Gamma(B_{s0})$ [keV]\\
\hline

Lu \textit{et al.} (Nonrelativistic) \cite{Luetal}
& $5756$ & $-0.90$ &$0.37$ &$0.70$ &$0.30$ &$34514$&$7.47$\\

Lakhina \textit{et al.} \cite{Lakhinaetal}
& $5776$ & $-0.89$ &$0.32$ &$0.65$ &$0.35$ &$34197$ & $10.42$\\

Ortega \textit{et al.} (Nonrelativistic) \cite{Ortegaetal}
& $5801$ & $-0.87$ & $0.28$ & $0.61$ & $0.39$ & $33927$ & $13.77$\\

Di Pierro \textit{et al.} \cite{DiPierroetal}
& $5804$ & $-0.87$ &$0.28$ &$0.61$ &$0.39$ &$33901$ &$14.14$\\

Godfrey \textit{et al.} ($2016$) \cite{Godfreyetal2016}
& $5805$ & $-0.87$ &$0.28$ &$0.61$ &$0.39$ &$33892$ &$14.27$ \\

Liu \textit{et al.} \cite{Liuetal}
& $5815$ & $-0.87$ &$0.27$ &$0.59$ &$0.41$ &$33812$ &$15.47$ \\

Sun \textit{et al.} \cite{Sunetal}
& $5830$ & $-0.86$ &$0.25$ &$0.58$ &$0.42$ & $33708$& $17.17$\\

Godfrey \textit{et al.} ($1991$) \cite{Godfreyetal1991}
& $5831$  &$-0.86$ &$0.25$ &$0.58$ &$0.42$ & $33701$& $17.28$\\

Ebert \textit{et al.} \cite{Ebertetal}
& $5833$ &$-0.86$ &$0.25$ &$0.57$ &$0.43$ &$33689$ &$17.50$\\

\hline 
\end{tabular} 
\end{table}

\begin{table}[htbp]
\centering
\caption{Same as in Table~\ref{tab:bare_cutoff1} but for $B_{s1}$.}
\label{tab:bare2_cutoff1}
\begin{tabular}{cccccccc}
\hline
Reference & $M_{\rm bare}$ [MeV] & $a_0$ [fm] & $r_0$ [fm] & $1-X$ & $X$ & $g_{\rm pole}$ [MeV] & $\Gamma(B_{s1}^{\ast})$ [keV] \\
\hline

Lu \textit{et al.} (Nonrelativistic) \cite{Luetal}
& $5801$ & $-0.90$ & $0.38$ & $0.70$ & $0.30$ & $34793$ & $2.24$ \\

Lakhina \textit{et al.} \cite{Lakhinaetal}
& $5803$ & $-0.90$ & $0.37$ & $0.69$ & $0.31$ & $34755$ & $2.33$ \\

Godfrey \textit{et al.} ($2016$) \cite{Godfrey:2016nwn}
& $5822$ & $-0.88$ & $0.32$ & $0.65$ & $0.35$ & $34459$ & $3.27$ \\

Di Pierro \textit{et al.} \cite{DiPierroetal}
& $5842$ & $-0.87$ & $0.29$ & $0.62$ & $0.38$ & $34236$ & $4.21$ \\

Liu \textit{et al.} \cite{Liuetal}
& $5843$ & $-0.87$ & $0.28$ & $0.62$ & $0.38$ & $34227$ & $4.25$ \\

Godfrey \textit{et al.} ($1991$) \cite{Godfreyetal1991}
& $5857$ & $-0.87$ & $0.27$ & $0.60$ & $0.40$ & $34106$ & $4.87$ \\

\begin{tabular}{c}
Ortega \textit{et al.} (Nonrelativistic) \cite{Ortegaetal} \\
Sun \textit{et al.} \cite{Sunetal}
\end{tabular}
& $5858$ & $-0.87$ & $0.27$ & $0.60$ & $0.40$ & $34099$ & $4.92$ \\

Ebert \textit{et al.} \cite{Ebertetal}
& $5865$ & $-0.86$ & $0.26$ & $0.59$ & $0.41$ & $34046$ & $5.21$ \\

\hline
\end{tabular}
\end{table}
\subsection{Off-shell coupled-channel calculation of the $B^{(*)}\bar{K}^{(*)}$ system}
In this subsection, we perform the off-shell coupled-channel calculation, including $S$ and $D$ partial waves, and allowing for the coupling between the $B^{(*)}\bar{K}^{(*)}$ channels. We solve the Lippmann-Schwinger equation after projecting the interaction into the specific partial waves. The interactions for the $B^{(*)}\bar{K}$ systems are given in Eqs.~(\ref{eq:rho}) and (\ref{eq:s1}). The formalism with the explicit form of the potentials for the $B^{(*)}\bar{K}^{(*)}$ coupled-channel systems is provided in~\cite{Shi:2024squ,Sanchez-Illana:2026cyv}. 
The difference here is in the vector-vector interaction, where we also add the contact term given by Eq.~(\ref{Eq:ctermL}). In the off-shell calculation, $t$ in Eq.~\eqref{eq:fullvmp} is not neglected. The hard cutoff $\Lambda$ in the integral Lippmann-Schwinger equation in Ref.~\cite{Sanchez-Illana:2026cyv} is adjusted to reproduce the observed state $B_{s0}$~\cite{LHCb:2026knl}. The results are presented in Table~\ref{Tab:pole_IIerr}, where a tentative assignment for the $B_{sJ}$ states in the PDG is shown. As can be seen, the decay width obtained for the state $J^P=0^+$ $B^*\bar{K}^*$ to $B\bar{K}$ is now compatible with that listed in the PDG for the $B_{sJ}(6114)$ in this decay mode, while that of the $J^P=2^+$ state is smaller, in line with the observed state $B_{sJ}(6063)$. See also~\cite{LHCb:2020pet}. While the results for the $B_{s0}$ and $B_{s1}$ states are similar to the ones of the on-shell single-channel scenario (Table~\ref{Tab:onshpole_Ierr}), the other poles have shifted masses. The largest difference comes from the $B^*\bar{K}^*$ $J^P=0^+, 1^+$ states, which now have masses of about $6130$~MeV, still below the $B^*\bar{K}^*$ threshold. The imaginary parts for the $B^*\bar{K}^*$ states now come from the contribution of the $B\bar{K}$ channel in the coupled-channel approach, which is open. The $B\bar{K}^*$ state can decay into $B^*\bar{K}$ with a decay width of around $14$~MeV. 

\begin{table}[htbp]
\caption{Pole positions calculated by the off-shell $T$-matrix. The cutoff are determined from the experimental mass of $B_{s0}(5700)$ with its value $\Lambda=1170\pm10$ MeV.}
\label{Tab:pole_IIerr}
\renewcommand{\arraystretch}{1.0}
\begin{tabular*}{\columnwidth}{@{\extracolsep\fill}lccccc}
\hline\hline 
 $J^{P}$ & Channel & Predicted state [MeV] & Experimental state [MeV] & RS & $E_b$ [MeV]\\[3pt]     
\hline

\multirow{4}{*}{$0^{+}$}
& \multirow{4}{*}{$B\bar K$--$B^\ast\bar K^\ast$}
& \multirow[c]{2}{*}{$M=5698.9\pm0.7$}
& \multirow[c]{2}{*}{\shortstack[c]{
$M=5698.9\pm1.5\pm0.6$\\[3pt]
$\Gamma<9.8(11.8)$
}}
& \multirow{2}{*}{$(+,+)$}
& \multirow{2}{*}{$-76.3\pm0.7$}\\[3pt]
&&&&&\\[3pt]

&& \multirow[c]{2}{*}{\shortstack[c]{
$M=6134.8\pm0.9$\\[3pt]
$\Gamma=40.2\pm0.6$
}}
& \multirow[c]{2}{*}{\shortstack[c]{
$M=6114\pm3\pm5$\\[3pt]
$\Gamma=66\pm18\pm21$
}}
& \multirow{2}{*}{$(-,+)$}
& \multirow{2}{*}{$-83.6\pm0.9$}\\
&&&&&\\

\hline

\multirow{4}{*}{$1^{+}$}
& \multirow{4}{*}{$B^\ast\bar K$--$B\bar K^\ast$}
& \multirow[c]{2}{*}{$M=5752.7\pm1.6$}
& \multirow[c]{2}{*}{--}
& \multirow{2}{*}{$(+,+)$}
& \multirow{2}{*}{$-67.7\pm1.6$}\\
&&&&&\\

&& \multirow[c]{2}{*}{\shortstack[c]{
$M=6022\pm2$\\[3pt]
$\Gamma=13.6\pm0.3$
}}
& \multirow[c]{2}{*}{--}
& \multirow{2}{*}{$(-,+)$}
& \multirow{2}{*}{$-151\pm2$}\\
&&&&&\\

\hline

\multirow{2}{*}{$1^+$}
& \multirow{2}{*}{$B^\ast\bar K^\ast$}
& \multirow[c]{2}{*}{\shortstack[c]{
$M=6136.0\pm0.9$\\[3pt]
$\Gamma=7.35\pm0.14$
}}
& \multirow[c]{2}{*}{--}
& \multirow{2}{*}{$(+)$}
& \multirow{2}{*}{$-82.4\pm0.9$}\\
&&&&&\\

\hline

\multirow{2}{*}{$2^{+}$}
& \multirow{2}{*}{$B\bar K$--$B^\ast\bar K^\ast$}
& \multirow[c]{2}{*}{\shortstack[c]{
$M=6060\pm2$\\
$\Gamma=5.29\pm0.03$
}}
& \multirow[c]{2}{*}{\shortstack[c]{
$M=6063.5\pm1.2\pm0.8$\\
$\Gamma=26\pm4\pm4$
}}
& \multirow{2}{*}{$(-,+)$}
& \multirow{2}{*}{$-158\pm2$}\\
&&&&&\\

\hline\hline
\end{tabular*}
\end{table}
\FloatBarrier
\section{Conclusions}\label{sec:sum}
In this work, we have performed a single-channel (on-shell) and coupled-channel (off-shell) calculation for the $B^{(*)}\bar{K}^{(*)}$ system within the hidden-gauge formalism. We find that for $I=0$ the interaction is attractive and leads to the generation of six states with spins $J^P=\lbrace 0,1,2\rbrace^+$. By adjusting the mass of the bound state obtained in the $B\bar K$ system to the new $B_{s0}(5700)$ state observed by the LHCb, we predict the masses of the other five states. In particular, we obtain three states with the quantum numbers $J^P=1^+$ near the $B^*\bar{K}$, $B\bar{K}^*$ and $B^*\bar{K}^*$ thresholds. Among them, the state lying close to the $B^*\bar{K}$ threshold, which is identified as the heavy-quark-spin partner of the $B_{s0}$, has a mass of around $5750$~MeV. Furthermore, we assign the $B_{sJ}(6063)$ and $B_{sJ}(6114)$ as the $2^+$ and $0^+$ $B^*\bar{K}^*$ molecular candidates, respectively, both of which decay into $B\bar{K}$ channel. In this assignment, the $J^P=2^+$ state has a smaller decay width than the $J^P=0^+$ state in our off-shell coupled-channel approach, in agreement with the experimental measurements. 

Regarding the $B_{s0}$ and $B_{s1}$ states, 
their favorable decay mode is the $B_s^{(*)0}\pi^0$, that can occur through isospin breaking process. We have considered two different cases. The one where the states are purely molecular, and the other where these can couple to a bare $b\bar{s}$ component. We have found that the coupling to this component can have a large impact on the decay width of these states. In particular, for the $B_{s0}$, this coupling can reduce the decay width from approximately $50$~keV to $10-20$~keV once considering the reasonable values of the bare mass $M_{\text{bare}}$. At present, the experimental data are not precise enough to determine this decay width, since only the upper limit of the total decay width is set to $\Gamma  <9.8(11.8)$~MeV~\cite{LHCb:2026knl}. We also have estimated that the $B_{s1}$ decay width is smaller than that of $B_{s0}$ because of heavy-quark-spin-violating contributions. From future precise measurements of these decay modes, the compositeness of these states can be determined. 

\begin{acknowledgements}

R.~Molina acknowledges support from the ESGENT program with Ref. ESGENT/018/2024 and the PROMETEU program with Ref. CIPROM/2023/59, of the Generalitat Valenciana, and also from the Spanish Ministerio de Economia y Competitividad and European Union (NextGenerationEU/PRTR) by the grant with Ref. CNS2022-136146, Ref. PID2023-147458NB-C21, and CEX2023-001292-S. This project has received funding from the European Union Horizon 2020 research and innovation program under the program H2020-INFRAIA-2018-1, grant agreement No. 824093 of the STRONG-2020 project.

\end{acknowledgements}

\appendix
\section{One loop integral for the $B_{s1}$ width}\label{AppA}
In this section we evaluate the one-loop integral related to the diagram in Fig. \ref{Fig:decay_diam} with pseudoscalar meson exchange that appears in the decay $B_{s1}\rightarrow B_s^{\ast 0}\pi^0$. The Feynman diagram with the notation used for the momentum is depicted in Fig. \ref{Fig:appA1} (right). For the triangle diagram in the left-hand side of Fig. \ref{Fig:appA1}, the solution reduces to the one calculated in~\cite{Gil-Dominguez:2023puj} for the $B_{s0}$ except for the product of polarization vectors incoming and outgoing, and of course, replacing the $B_{(s)}$ by the $B^*_{(s)}$.

\begin{figure}[h]
\centering
\includegraphics[height=0.25\columnwidth]{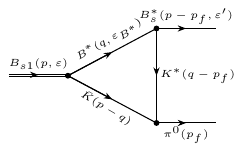}
\hspace{0.02\columnwidth}
\includegraphics[height=0.25\columnwidth]{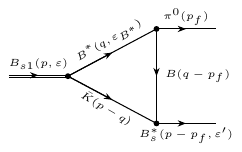}
\caption{Feynman diagrams for the decay $B_{s1}\to B^*_{s}\pi^0$.}
\label{Fig:appA1}
\end{figure}
The interaction for every vertex involved on the right side of Fig. \ref{Fig:appA1} is:
\begin{align}
    -it_{B_{s1},B^\ast \bar K} = -ig_{B_{s1},B^\ast \bar K}(\varepsilon \cdot\varepsilon_{B^\ast}),\nonumber\\
    -it_{B^\ast,B \pi} = -ia\varepsilon_{B^\ast}\cdot(2p_f-q),\nonumber\\-it_{\bar K B,B_s^\ast} = -ib\varepsilon^{\prime\ast} \cdot(p-2q+p_f).
    \label{vertexApp1}
\end{align}
In the approximation of low momentum compared with the mass of the heavy vector meson, we can write the loop integral in Fig. \ref{Fig:appA1} as
\begin{equation}
    -it_{22}=F\int \frac{{d}^4q}{(2\pi)^4}\frac{(2\vec p_f-\vec q)\cdot \varepsilon(\vec p_f -2\vec q)\cdot \varepsilon^{\prime\ast}}{(q^2-m_{B^\ast}^2+i\epsilon)\left((q-p_f)^2-m_B^2+i\epsilon\right)\left((p-q)^2-m_K^2+i\epsilon\right)},
    \label{App1loop}
\end{equation}
with $F=ia\cdot b\cdot g_{B_{s1},B^\ast\bar K}$, being $a$ and $b$ the coefficients from the Hidden Gauge vertices, and the denominator $D\equiv(q^2-m_{B^\ast}^2+i\epsilon)\left((q-p_f)^2-m_B^2+i\epsilon\right)\left((p-q)^2-m_K^2+i\epsilon\right)$. This integral has a different structure than the integral coming from the left side of Fig. \ref{Fig:appA1} where the product of polarization vectors $\varepsilon \cdot\varepsilon^{\ast \prime}$ is factored out. This integral can be recast as
\begin{equation}
    -it_{22}=F\varepsilon^i\varepsilon^{\ast j}\left[Ap_f^ip_f^j+B\delta^{ij}\right],
    \label{recast}
\end{equation}
with
\begin{align}
    A=\int \frac{d^4q}{(2\pi)^4}\frac{f_A(\vec q, \vec p_f)}{D},\qquad
    B=\int \frac{d^4q}{(2\pi)^4}\frac{f_B(\vec q, \vec p_f)}{D},
    \label{AppAyB}
\end{align}
being
\begin{align}
    f_A(\vec q, \vec p_f)&=2+\frac{1}{|\vec p_f|^2}\left[\frac{3(\vec p_f\cdot \vec q)^2}{|\vec p_f|^2}-\vec q^{\hspace{0.1cm}2}-5\vec q\cdot\vec p_f\right],\nonumber\\
    f_B(\vec q, \vec p_f)&=\vec q^{\hspace{0.1cm}2}-\frac{(\vec p_f\cdot \vec q)^2}{|\vec p_f|^2}.
    \label{Appf}
\end{align}
We need to evaluate an integral like 
\begin{equation}
    I_2=\int \frac{d^4q}{(2\pi)^4}\frac{f(\vec q, \vec p_f)}{D}.
    \label{AppI2}
\end{equation}
Thus, we decompose the propagator into two parts,
\begin{align}
    \frac{1}{q^2-m_{B^\ast}^2+i\epsilon}&=\frac{1}{2\omega_{B^\ast}}\left(\frac{1}{q^0-\omega_{B^\ast}+i\epsilon}-\frac{1}{q^0+\omega_{B^\ast}-i\epsilon}\right),\nonumber\\
    \frac{1}{(q-p_f)^2-m_B^2+i\epsilon}&=\frac{1}{2\omega_B}\left(\frac{1}{(q^0-p_f^0)-\omega_B+i\epsilon}-\frac{1}{(q^0-p_f^0)+\omega_B-i\epsilon}\right).
    \label{Appdecomp}
\end{align}
For the $B^{(\ast)}$ meson we can neglect the negative energy part of the propagator. Hence, the simplified integral is 
\begin{equation}
    I_2=\int\frac{d^4q}{(2\pi)^4}\frac{f(\vec q, \vec p_f)}{4\omega_B\omega_{B^\ast}((p-q)^2-m_K^2+i\epsilon)(q^0-\omega_{B^\ast}+i\epsilon)(q^0-p_f^0-\omega_B+i\epsilon)},
    \label{Appsimpint}
\end{equation}
with $\omega_K=\sqrt{\vec{q}^{\hspace{0.1cm}2}+m_K^2}$, $\omega_B=\sqrt{(\vec q-\vec p_f)^2+m_B^2}$ and $\omega_{B^\ast}=\sqrt{\vec q^{\hspace{0.1cm}2}+m_{B^\ast}^2}$.

By performing the integral through the Residue Theorem, we obtain
\begin{equation}
    I_2=-i\int\frac{d\,{q}\,d\,\text{cos}\,\theta}{(2\pi)^2}\frac{|\vec q|^2 f(\vec q, \vec p_f)}{8\omega_K\omega_B\omega_{B^\ast}(p^0-\omega_K-\omega_{B^\ast}+i\epsilon)(p^0-p_f^0-\omega_K-\omega_B+i\epsilon)}.
    \label{Appsimpint2}
\end{equation}
Finally, we need to sum the amplitude from the two top diagrams in Fig. \ref{Fig:decay_diam}, depicted on the left side of Fig. \ref{Fig:appA1}, of the type
\begin{equation}
    -it_{21}=\frac{ig_{B_{s1},B^\ast\bar K} g^2 c_1}{32\pi^2}I_1(\omega_{B^\ast}, \omega_{\bar K}, \omega_{K^\ast})\vec \varepsilon\cdot\vec \varepsilon^{\hspace{0.1cm}\prime\ast},
    \label{AppFGil}
\end{equation}
where $\omega_{K^\ast}=\sqrt{(\vec q -\vec p_f)^2+m_{K^\ast}^2}$.
The function $I_1$ was calculated in \cite{Gil-Dominguez:2023puj},
\begin{align}  I_1(\omega_1,\omega_2,\omega_M)=&\int_{0}^{\Lambda}\frac{d^3q}{\omega_{1}\omega_2\omega_{M}}\left[\frac{(p^0)^2-(p_f^0-p^0+\omega_2)^2+(\vec p_f-\vec q)^2}{(p^0-\omega_{1}-\omega_2+i\epsilon)(p_f^0-\omega_{M}-\omega_2)}+\frac{1}{p^0-p_f^0-\omega_{1}-\omega_{M}+i\epsilon}\right.\nonumber\\
& \left.\times\left(\frac{(p^0)^2-(p_f^0-\omega_{1})^2+(\vec p_f-\vec q)^2}{p^0-\omega_{1}-\omega_2+i\epsilon}-\frac{(p^0)^2-(2p_f^0-p^0+\omega_{M})^2+(\vec p_f-\vec q)^2}{p_f^0+\omega_{M}+\omega_2}\right)\right],
\label{AppI1}
\end{align}
    %
Hence, we get
\begin{equation}
    -it_2=-i(t_{21}+t_{22})=F\cdot A(\vec p_f\cdot \vec \varepsilon) (\vec p_f\cdot \vec \varepsilon^{\hspace{0.1cm}\prime\ast})+\left[\frac{ig_{B_{s1},B^\ast\bar K}g^2 c_1}{32\pi^2} I_1(\omega_{B^\ast}, \omega_{\bar K}, \omega_{K^\ast})+B\cdot F\right](\vec \varepsilon\cdot \vec \varepsilon^{\hspace{0.1cm}\prime\ast}),
    \label{Appsumt}
\end{equation}
and we define $C\equiv ig_{B_{s1},B^\ast\bar K}g^2 c_1 I_1(\omega_{B^\ast}, \omega_{\bar K}, \omega_{K^\ast})/32\pi^2+B\cdot F$. After averaging over initial polarizations and summing over final polarizations, we obtain
\begin{equation}
    \sum_{\lambda_i,\lambda_f} |t_2|^2=|\vec p_f|^2\left[|F\cdot A|^2|\vec p_f|^2+
2\,\mathrm{Re}\left(F\cdot A\cdot C^*\right)+3\frac{|C|^2}{|\vec p_f|^2}\right].
\label{Appsumsum}
\end{equation}

\section{Gaussian regulator}\label{AppB}
In this section, we present the results using a Gaussian regulator in the green function (Eq.~\ref{Eq:greenG}), instead of the hard cutoff shown in Eq.~\ref{Eq:green}. With the parameter $\Lambda$ between $500$ MeV and $800$ MeV we find the results in Table~\ref{tab:bare_cutoff2} for $B_{s0}$ and Table~\ref{tab:bare_cutoff3} for $B_{s1}$ compatible with the values given using a hard cutoff in Table~\ref{tab:bare_cutoff1} for $B_{s0}$ and Table~\ref{tab:bare2_cutoff1} for $B_{s1}$.
\begin{align}
G(s)=\int_{0}^{\infty}\frac{d^3{\boldsymbol q}}{(2\pi)^3}\frac{\omega_1(\boldsymbol q)+\omega_2(\boldsymbol q)}{2\omega_1(\boldsymbol q)\omega_2(\boldsymbol q)}\frac{e^{-2(\boldsymbol{q}^2-\boldsymbol{p}^2)/\Lambda^2}}{s-\left[\omega_1(\boldsymbol q)+\omega_2(\boldsymbol q)\right]^2+i\epsilon}.
\label{Eq:greenG}
\end{align}
where $\boldsymbol p$ is the three-momentum in the center-of-mass (c.m.) frame.

\begin{table}[htbp] \centering
\caption{Values of the bare mass $M_{\rm bare}$, the corresponding scattering length $a_0$, effective range $r_0$, compositeness $X$, complementary component $1-X$, pole coupling $g_{\rm pole}$, and decay width $\Gamma(B_{s0})$. The compositeness and pole coupling are extracted from the $T$-matrix regularized by the Gaussian form factor in Eq.~\eqref{Eq:greenG}.}
\label{tab:bare_cutoff2}
\begin{tabular}{cccccccc} 
\hline
Reference & $M_{\rm bare}$ [MeV] & $a_0$ [fm] & $r_0$ [fm] & $1-X$ & $X$ & $g_{\rm pole}$ [MeV] & $\Gamma(B_{s0})$ [keV]\\
\hline

Lu \textit{et al.} (Nonrelativistic) \cite{Luetal}
& $5756$ & $-0.83$ & $0.16$ & $0.63$ & $0.37$ & $32856$ & $8.39$ \\ 

Lakhina \textit{et al.} \cite{Lakhinaetal}
& $5776$ & $-0.82$ & $0.14$ & $0.60$ & $0.40$ & $32616$ & $10.97$ \\ 

Ortega \textit{et al.} (Nonrelativistic) \cite{Ortegaetal}
& $5801$ & $-0.81$ & $0.13$ & $0.56$ & $0.44$ & $32438$ & $13.80$ \\ 

Di Pierro \textit{et al.} \cite{DiPierroetal}
& $5804$ & $-0.81$ & $0.13$ & $0.56$ & $0.44$ & $32422$ & $14.12$ \\ 

Godfrey \textit{et al.} ($2016$) \cite{Godfreyetal2016}
& $5805$ & $-0.81$ & $0.13$ & $0.56$ & $0.44$ & $32417$ & $14.22$ \\ 

Liu \textit{et al.} \cite{Liuetal}
& $5815$ & $-0.81$ & $0.12$ & $0.54$ & $0.46$ & $32370$ & $15.22$ \\ 

Sun \textit{et al.} \cite{Sunetal}
& $5830$ & $-0.81$ & $0.12$ & $0.53$ & $0.46$ & $32311$ & $16.61$ \\ 

Godfrey \textit{et al.} ($1991$) \cite{Godfreyetal1991}
& $5831$ & $-0.81$ & $0.12$ & $0.52$ & $0.47$ & $32308$ & $16.70$ \\ 

Ebert \textit{et al.} \cite{Ebertetal}
& $5833$ & $-0.81$ & $0.12$ & $0.53$ & $0.47$ & $32301$ & $16.88$ \\ 

\hline
\end{tabular} 
\end{table}

\begin{table}[htbp]
\centering
\caption{Same as in Table~\ref{tab:bare_cutoff2} but for $B_{s1}$.}
\label{tab:bare_cutoff3}
\begin{tabular}{cccccccc}
\hline
Reference & $M_{\rm bare}$ [MeV] & $a_0$ [fm] & $r_0$ [fm] & $1-X$ & $X$ & $g_{\rm pole}$ [MeV] & $\Gamma(B_{s1}^{\ast})$ [keV] \\
\hline

Lu \textit{et al.} (Nonrelativistic) \cite{Luetal}
& $5801$ & $-0.82$ & $0.16$ & $0.63$ & $0.37$ & $33112$ & $2.24$ \\

Lakhina \textit{et al.} \cite{Lakhinaetal}
& $5803$ & $-0.82$ & $0.15$ & $0.63$ & $0.37$ & $33081$ & $2.33$ \\

Godfrey \textit{et al.} ($2016$) \cite{Godfreyetal2016}
& $5822$ & $-0.82$ & $0.14$ & $0.59$ & $0.41$ & $32861$ & $3.12$ \\

Di Pierro \textit{et al.} \cite{DiPierroetal}
& $5842$ & $-0.81$ & $0.13$ & $0.57$ & $0.43$ & $32714$ & $3.89$ \\

Liu \textit{et al.} \cite{Liuetal}
& $5843$ & $-0.81$ & $0.13$ & $0.56$ & $0.44$ & $32708$ & $3.93$ \\

Godfrey \textit{et al.} ($1991$) \cite{Godfreyetal1991}
& $5857$ & $-0.81$ & $0.12$ & $0.55$ & $0.45$ & $32636$ & $4.44$ \\

\begin{tabular}{c}
Ortega \textit{et al.} (Nonrelativistic) \cite{Ortegaetal} \\
Sun \textit{et al.} \cite{Sunetal}
\end{tabular}
& $5858$ & $-0.81$ & $0.12$ & $0.55$ & $0.45$ & $32632$ & $4.47$ \\

Ebert \textit{et al.} \cite{Ebertetal}
& $5865$ & $-0.81$ & $0.12$ & $0.54$ & $0.46$ & $32602$ & $4.72$ \\

\hline
\end{tabular}
\end{table}

\newpage
\bibliography{ref}
\end{document}